# New-Generation Design-Technology Co-Optimization (DTCO): Machine-Learning Assisted Modeling Framework

Zhe Zhang[1], Runsheng Wang[1*], Cheng Chen[1], Qianqian Huang[1], Yangyuan Wang[1],
Cheng Hu[2], Dehuang Wu[2], Joddy Wang[2], Ru Huang[1]
[1]Institute of Microelectronics, Peking University, Beijing 100871, China
[2]Synopsys, Inc., Mountain View, CA 94043, USA
*Email: r.wang@pku.edu.cn

*Abstract* — In this paper, we propose a machine-learning assisted modeling framework in design-technology co-optimization (DTCO) flow. Neural network (NN) based surrogate model is used as an alternative of compact model of new devices without prior knowledge of device physics to predict device and circuit electrical characteristics. This modeling framework is demonstrated and verified in FinFET with high predicted accuracy in device and circuit level. Details about the data handling and prediction results are discussed. Moreover, same framework is applied to new mechanism device tunnel FET (TFET) to predict device and circuit characteristics. This work provides new modeling method for DTCO flow.

## I. Introduction

With device continuously scaling, new structure, material and mechanism devices are proposed to meet different performance constrains in real life [1-3]. Rapid technology development requires new design framework. A methodology named design-technology co-optimization (DTCO) is proposed to reduce cost and time-to-market in advanced process development (Fig. 1) [4]. SPICE model plays a key role in the forward technology-to-design flow. For conventional device structure, SPICE compact model is slightly modified in different technology generations. However, for devices with new mechanisms, such as TFET [2-3] and negative capacitance FET [1], the underlying physics is not fully clear in the early device development stage. Complex physics is difficult to be abstracted as formula in compact model. To evaluate the circuit performance of new devices in early stage, a data-oriented surrogate model is urgently needed to catch up with rapid technology development. Machine learning algorithm such as NN can compute nonlinear equations for multivariate inputs and imitate the complex physical equations in real device [5-6]. In this paper, NN is used as a surrogate model to evaluate FinFET and TFET device and circuit characteristics in forward DTCO flow.

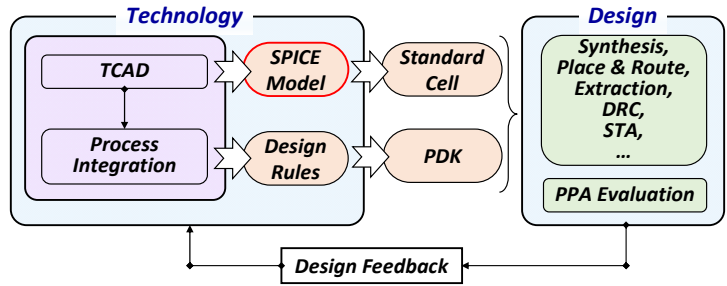

Fig. 1 Simplified design-technology co-optimization (DTCO) flow for new devices.

## II. Demonstrations and Verifications in FinFET

Precise compact model is a gold standard for circuit simulation. Traditional circuit simulation uses mature compact model (e.g., BSIM-CMG) to evaluate circuit performance. In NN based surrogate model (Fig.2), the weights and biases are trained from TCAD results or silicon data. Inputs can be bias conditions, geometry information. Output results such as I, G, Q, C are essential for further circuit simulation.

To compare the differences between traditional simulation method and proposed surrogate model, first, the advanced 16/14nm FinFET device is adopted for demonstrations due to its complete BSIM-CMG compact model. In this part, the data for NN training comes from SPICE results with inputs set as $V_g$, $V_d$, $V_s$ ranging from 0 to 0.8V with interval equals to 50mV. The simulated data when $V_d>=V_s$ in Fig. 3 show different distributions, which means data pre-processing is important before training. The results for $V_d<V_s$ and $V_d>V_s$ have symmetry value. Only three-terminal results are enough for FinFET. Other electrical characteristics can be deduced from current and charge information, which reduces the complexity in NN training. Fig. 4 shows the training loss with different neurons and layers. For a multi-input and multi-output regression model in this case, the proper neurons, layers, activation function, and optimizer should be carefully considered. After fine-tuning, the predicted relative error for I and Q (nFinFET) are shown in Fig. 5 with very small mean value shown in the inset. Same procedure is performed for pFinFET under full bias conditions. The trained NN integrated with HSPICE [7] is carried out for SRAM circuit simulation. Fig. 6 shows the butterfly curves and N-curves of read and hold process from direct SPICE results and surrogate model results, which show great consistency.

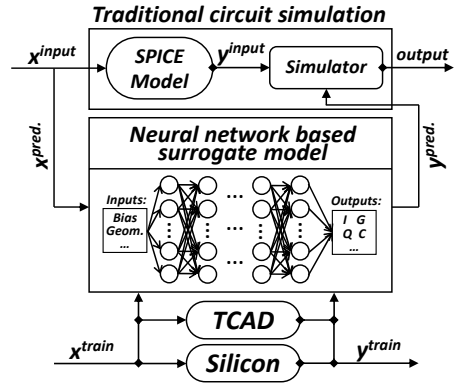

Fig. 2 Neural network (NN) based surrogate model for simulations.

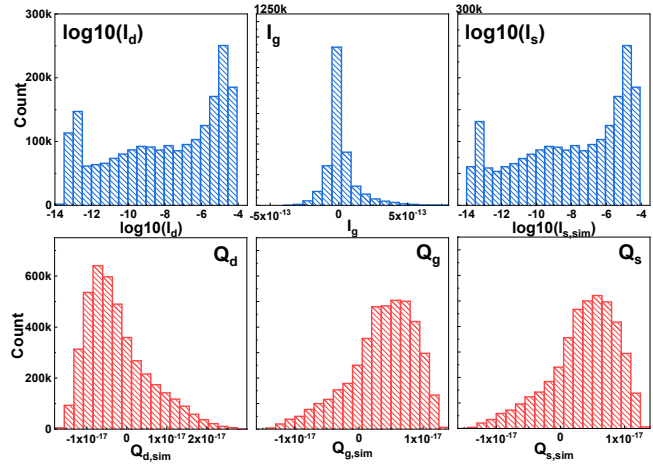

Fig. 3 Histogram of current and charge of three terminals (drain, gate, source) in nFinFET when $V_d >= V_s$. Results are simulated from HSPICE.

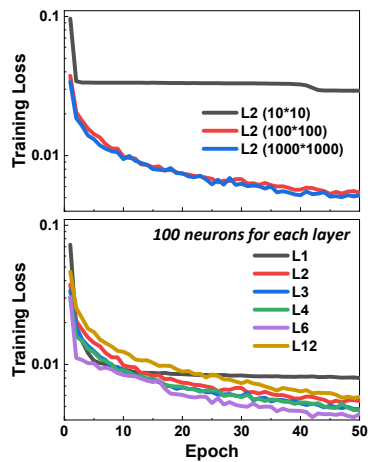

Fig. 4 Training of multilayer neural networks on device electrical characteristics. (a) Two layers with different neurons. (b) Fixed neurons with different layers.

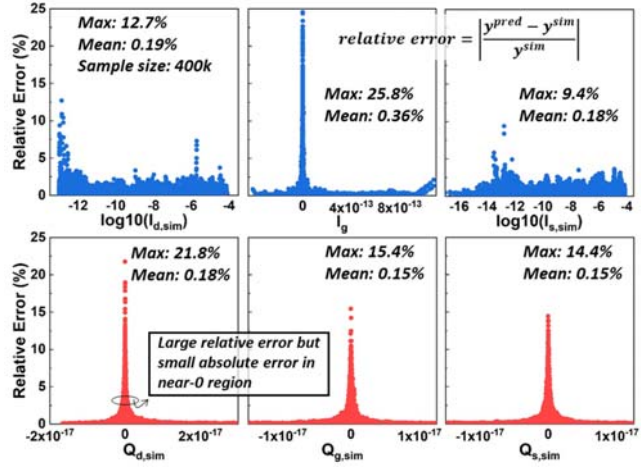

Fig. 5 Scatter plots of relative error of predicted results and simulated results. 400k test data different from training data show small mean relative error.

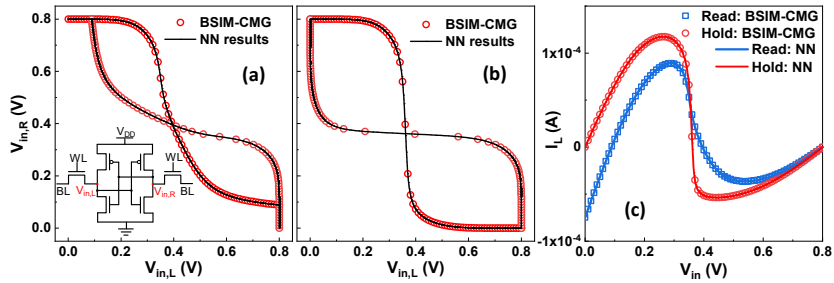

Fig. 6 Butterfly curves of (a) read and (b) hold process in FinFET SRAM [inset in (a)]. (c) N-curves of read and hold process. Results of NN model match well with direct simulated results based on BSIM-CMG.

III. PREDICTIONS AND RESULTS IN TUNNEL FET

TFET has ultra-steep subthreshold slop and is supposed to be a promising candidate for ultra-low power logic circuit. Different from FinFET, the physical mechanism of current generation in TFET is mainly band-to-band tunneling. That makes it difficult for existing planar device or FinFET compact model extended to TFET with minor modifications. Developing a new physics-based model for TFET is time-consuming in the early process stage. A surrogate model, which no device physics is needed, can help to imitate electrical characteristics and perform circuit simulation and thus accelerating the DTCO flow.

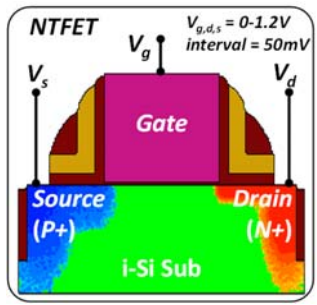

Fig. 7 2D N-type TFET (nTFET) structure used for training data collection in TCAD simulations.

The data source for training in TFET comes from TCAD [8]. Fig. 7 is 2D schematic view of a simulated N-type TFET (nTFET) with p-i-n structure. Fig. 8 shows the transfer curves of nTFET under different $V_d$ in linear and logarithmic region. The reverse drain current of nTFET is forward p-n junction current due to the p-i-n structure. Therefore, the current (Fig. 9) and other electrical characteristics under $V_d>V_s$ and $V_d<V_s$ have different distributions and should be trained separately for better precision. Coefficient of determination is adopted to benchmark the accuracy. More training data can lead to more accurate predictions (Fig. 10), but the time for data collection from TCAD is longer. Fig. 11 shows the scatter plot of predicted drain current and simulated drain current. Details in the above/near-$V_{th}$ region, which is more important for circuit performance evaluation, is enlarged in Fig. 11(b). The predicted mean relative error is small in all bias conditions (Fig. 12). Transfer curves of predicted results in Fig. 13 exhibit a good agreement with simulated results and show smooth and continuous characteristics.

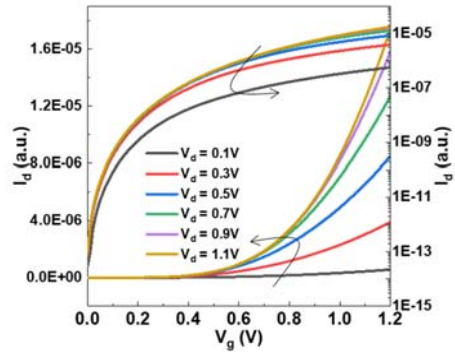

Fig. 8 Simulated transfer curves of nTFET under different $V_d$ in linear and logarithmic scale.

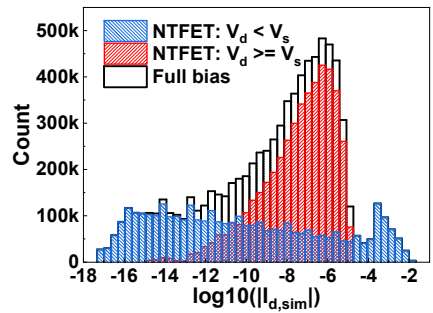

Fig. 9 Histogram of current of drain terminal in nTFET. Distributions are different in different bias conditions.

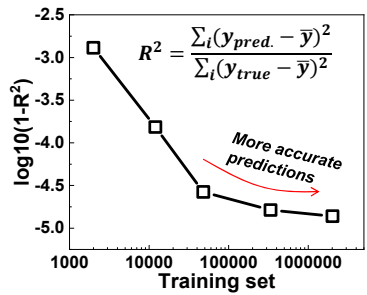

Fig. 10 Coefficient of determination ($R^2$) for different size of training set.

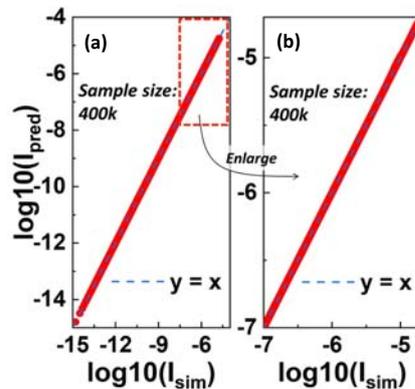

Fig. 11 Scatter plots of predicted $I_d$ and simulated $I_d$ in nTFET under $V_d >= V_s$ for (a) full scale; (b) above/near $V_{th}$ region.

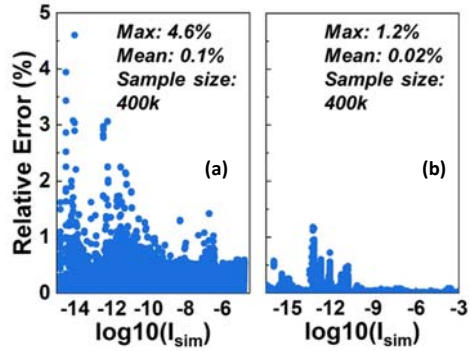

Fig. 12 Scatter plots of predicted relative error and simulated results in nTFET when (a) $V_d >= V_s$ and (b) $V_d < V_s$.

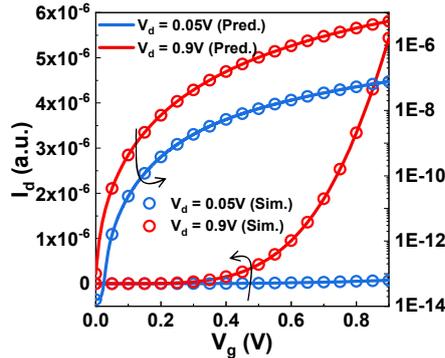

Fig. 13 Transfer curves of predicted results and simulated results under $V_d$ = 50mV, 0.9V.

Both nTFET and pTFET NN models are trained and fine-tuned for next-step circuit simulations. TFET inverter is introduced to predict circuit DC performance, as shown in Fig. 14 for demonstration. As for transient simulation, a 2-NAND TFET circuit is adopted in Fig. 15. The first-stage output voltage shows coupling capacitance noise due to large $C_{gd}$ and unidirectional conduction in TFET device when $V_{in,2}$ changes from '0' to '1' [9]. This abnormal characteristic can also be realized by the surrogate models, which proves the predictability of the models.

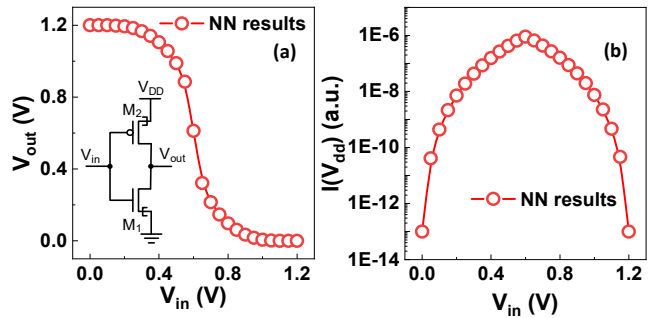

Fig. 14 TFET circuit simulation using NN based surrogate model. (a) DC simulation results of a TFET inverter. (b) Current in $V_{DD}$.

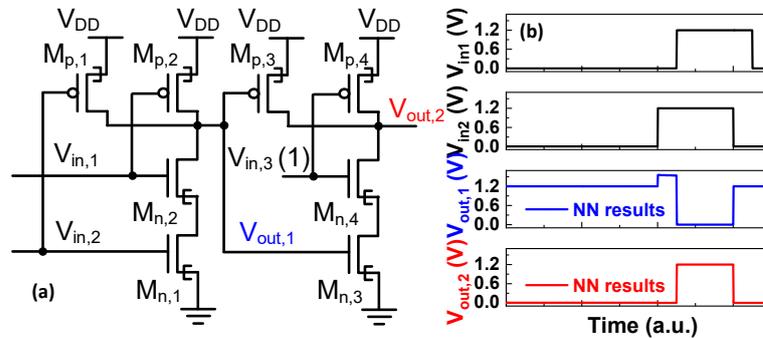

Fig. 15 TFET circuit simulation using NN based surrogate model. (a) 2-NAND structure. (b) Transient simulation results of first and second stage voltage outputs.

## IV. Conclusions

For rapid technology evolution, a surrogate NN model can replace the compact model of new mechanism device to speed up DTCO flow. The high predicted accuracy is achieved in both device and circuit level of FinFET and TFET. This machine-learning assisted modeling framework can be an alternative component for current DTCO flow.

## Acknowledgments

This work was support by NSFC (61522402 and 61421005) and the 111 project (B18001).